%% file: main.tex
\documentclass[sigconf, screen]{acmart}

\input{preamble/packages}
\input{preamble/definitions}
\input{preamble/authors}
\input{preamble/metadata}

\begin{document}

\title{An Empirical Study of Evaluating Long-form Question Answering}

%\thanks{\textsuperscript{\dag}Both authors contributed equally to this research.}
%\thanks{\textsuperscript{*}Corresponding author.}

\begin{abstract}
\Ac{LFQA} aims to generate lengthy answers to complex questions. This scenario presents great flexibility as well as significant challenges for evaluation. 
Most evaluations rely on deterministic metrics that depend on string or n-gram matching, while the reliability of large language model-based evaluations for long-form answers remains relatively unexplored.
We address this gap by conducting an in-depth study of long-form answer evaluation with the following research questions: (i) To what extent do existing automatic evaluation metrics serve as a substitute for human evaluations? (ii) What are the limitations of existing evaluation metrics compared to human evaluations? (iii) How can the effectiveness and robustness of existing evaluation methods be improved?  
We collect 5,236 factoid and non-factoid long-form answers generated by different large language models and conduct a human evaluation on 2,079 of them, focusing on correctness and informativeness. Subsequently, we investigated the performance of automatic evaluation metrics by evaluating these answers, analyzing the consistency between these metrics and human evaluations. We find that the style, length of the answers, and the category of questions can bias the automatic evaluation metrics. However, fine-grained evaluation helps mitigate this issue on some metrics.
Our findings have important implications for the use of large language models for evaluating long-form question answering.
All code and datasets are available at \url{https://github.com/bugtig6351/lfqa_evaluation}.
% We release all our human evaluation data, annotation interface and code to facilitate future research.\footnote{\url{https://github.com/bugtig6351/lfqa_evaluation}}
\end{abstract}

\maketitle

\vspace{-12pt}
%\acresetall

\section{Introduction}

\Ac{LFQA} aims to enable generation or retrieval models to answer open-ended questions with long answers at the paragraph level \cite{Fan2019eli5, Krishna2021Hurdles}. With the advancement of \acp{LLM}, the ability to generate long-form answers has improved significantly \cite{mann2020language_gpt3, bubeck2023sparks_gpt4}. While such long-form answers could address more complex and diverse questions, their flexibility presents significant challenges for evaluation. 

Currently, much of the research on \ac{LFQA} focuses on designing models and frameworks to address the hallucination of \ac{LLM} and improve overall performance. 
For example, \citet{su2022read_before_generate} propose a framework that uses fine-grained, answer-related salient information to enhance the faithfulness of model responses and reduce hallucinations. 
%Chen et al.\cite{chen2023understanding_rag} analyzed how different qualities of retrieved document sets affect the quality of model responses. 
\citet{tao2024chainofdiscussion} introduce a chain-of-discussion framework that takes advantage of synergy among multiple open source \acp{LLM} to provide more accurate and comprehensive answers. 
Despite these advancements, most studies still rely on simple \ac{QA} metrics, such as ROUGE~\cite{lin2004rouge} and Exact Match\cite{Stelmakh2022asqa}, which are based on string or n-gram matching, to evaluate long-form answers. However, these metrics often fail to capture the nuanced and flexible evaluation required by the complex structure of long-form answers, demonstrating weak correlation with human judgment \cite{Krishna2021Hurdles, liu2023geval}.

\heading{LLM-based evaluators}
Recently, there has been growing interest in developing LLM-based evaluators to provide more reliable assessments of long-form answers \cite{chiang2023closer}. 
For example, \citet{vu2024foundational} introduce LLM autoraters by training the PaLM-2-24B model on a collection of 102 quality assessment tasks comprising more than 5.3M human judgements (i.e., FLAMe). 
% Sewon et al.~\cite{min2023factscore} proposed FactScore to evaluate the factual precision by breaking the result into a series of atomic facts. 
\citet{fan2024eva} propose an LLM-based metric, EVA-Score, to evaluate the informativeness in abstract lonng-form summarization.
% G-Eval or ...
Although \acp{LLM} have shown promise in providing more comprehensive evaluations of long-form answers, they remain prone to vulnerabilities due to their well-known hallucination issues \cite{wang2023chatgpt_evaluator}.

\heading{Goals and questions}
% In this paper,  we focus on the performance of automatic metrics in LFQA tasks, including traditional methods and LLM-based methods, and compare them with human evaluators. We compare the correlation between human and automatic metrics themselves and examine potential limitations of different evaluators, including biases related to length and type of questions. 
In this paper, we aim to fill the above gap by conducting a comprehensive analysis of existing automatic evaluation metrics, including deterministic metrics and model-based metrics. We compare automatic metrics with human evaluations to examine their strengths and limitations.
Given this setup, we explore the following three research questions:
\begin{enumerate}[label=(RQ\arabic*),leftmargin=*]
    \item To what extent do existing automatic evaluation metrics serve as substitutes for human evaluations in \ac{LFQA}, and how accurate are these metrics?

    \item What are the limitations of automatic evaluation metrics compared to human evaluators? What influences their stability and fairness?

    \item How can the effectiveness and robustness of existing evaluation methods be improved?
    % What adjustments can we make to the evaluation metrics to avoid these biases and limitations?
\end{enumerate}

\heading{Main findings}
Our study involves both factual and non-factual \ac{LFQA} tasks, collecting answers generated by seven different \acp{LLM} on the ASQA, ANTIQUE and Wikieval datasets, and performing a human evaluation on a subset of these answers (over 2,079 answer pairs with 4,158 ratings and justifications). 

%Our findings can be summarized as follows.  
\begin{itemize}[leftmargin=*]
    \item For RQ1, we find that metrics based on large models demonstrate significantly higher consistency with human evaluations than deterministic metrics. They are also more stable in assessing different types of \ac{LFQA}. 
    %Some deterministic metrics showed almost zero consistency with human judgments.
    \item For RQ2, we conduct an analysis from two perspectives. First, we examine the impact of meaningless minor perturbations to the prompt on evaluation results, assessing the stability of the evaluators. Second, we investigate whether factors such as text length, question type, and non-semantic variations in phrasing introduce evaluator bias, thereby affecting the fairness of the evaluators. Our findings indicate that deterministic metrics are influenced by the length of the reference text and tend to penalize longer answers. In contrast, LLM-based methods do not exhibit any significant biases in this regard.
    % \item For RQ3, our analysis reveals the tendency of metrics within the same question category to exhibit similar error patterns during evaluation, highlighting the superior accuracy of fine-grained assessment methods over their counterparts.
    \item For RQ3, our research analyzes the impact of different prompting strategies on outcomes, highlighting that fine-grained evaluation methods exhibit higher accuracy compared to similar evaluation approaches.
\end{itemize}

% \noindent%

\section{Related Work}

This work aims to analyze the shortcomings of existing evaluations in LFQA and provide guidance for future efforts. Therefore, we review previous work, including (i) automated evaluation methods, and (ii) evaluation of evaluators.

\subsection{Automatic Evaluation Methods}

To evaluate the performance of large models, numerous studies have used automatic evaluation metrics across multiple independent benchmarks \cite{helm_2022, Systematic_Study_2023, Answerability_2023}. These metrics are generally categorized into three types: n-gram-based, embedding-based, and LLM-based. Each category offers distinct approaches for assessing model responses, aiming to balance efficiency and accuracy in evaluating various aspects of generated text.

N-gram-based metrics, such as ROUGE \cite{lin2004rouge} and BLEU \cite{papineni2002bleu}, are widely adopted due to their high efficiency and low cost, primarily verifying the correctness of model responses. However, these metrics can be adversely affected by syntactic errors \cite{Reiter_Belz_2009} and may struggle to capture human preferences when comparing outputs from different models \cite{Graham_Translationese_2019, Edunov_eval_trans_2020}, with recent studies further supporting these limitations \cite{Krishna2021Hurdles, xu2023critical}. Embedding-based methods, such as BERTScore \cite{zhang2019bertscore} and MoverScore \cite{Zhao_2019_moverscore}, use pretrained language models to better capture semantic similarities, thereby reducing the impact of superficial textual changes. Despite their advancements, these methods rely heavily on high-quality reference texts, posing challenges for open-ended questions where crafting appropriate references is difficult \cite{Chen_2023ref_free}.

While embedding-based approaches improve upon traditional n-gram methods by focusing on semantic content, their dependence on reference texts limits their applicability in more flexible evaluation scenarios. On the other hand, LLM-based evaluators offer a novel approach for reference-free evaluation, demonstrating reasonable performance in zero-shot settings \cite{Aiyappa_An_Kwak_Ahn_2023_chatgpt_eval, Chiang_Lee_2023_Alternative_eval, Liusie_Manakul_Gales_2023_zero_shot}. Examples include GPTScore \cite{fu2023gptscore}, G-EVAL \cite{liu2023geval}, LLM-EVAL \cite{lin2023llm_eval}, and RAGAS \cite{es2023ragas}, which leverage large language models to provide multi-dimensional evaluations. However, LLMs may inherit biases from their training data \cite{wang2023llm_not_fair} and generate hallucinations \cite{Ji_2023_hallucination}, with studies confirming biases in these evaluators \cite{wang2023llm_not_fair, zheng2024judging, wang2023chatgpt_evaluator}. Although fine-grained evaluations can help mitigate some issues \cite{Min_2023_factscore, Ye_2023_flask}, the overall reliability of automatic evaluation methods as substitutes for human evaluators remains an area requiring further investigation.

\subsection{Evaluation of Evaluators}

Since LLM-as-a-Judge has become a new evaluation paradigm, many studies have assessed the effectiveness of these methods. The basic approach of previous research has been to treat various types of evaluators as virtual annotators and evaluate their consistency with human annotators to achieve alignment with human\cite{llm_judge_survey_2024}.

MT-bench and the Chatbot arena dataset~\cite{zheng2024judging} contain a small amount of human-crafted queries annotated by experts and a large volume of crowdsourced user preferences data from real-world users, analyzing the agreement between LLM evaluators and human annotators and bias of the evaluator. \citet{xu2023critical} collect the consistency between automatic metrics and human experts in the evaluation of LFQA across seven different knowledge domains, focus on overall answer preference, coherence, and factuality. PandaLM~\cite{pandaLM_2024} trains a LLM evaluator and constructed a general instruction-tuning benchmark that includes a significant amount of automated scoring and human preference data, using a partial order graph to compare the performance of the models. 

In this work, we will adopt previous evaluation approaches for meta-evaluation, assessing the performance of evaluators based on their consistency with human annotators. We will focus on the biases and robustness exhibited by different evaluation methods throughout this process.

\section{A Study of \ac{LFQA} Evaluation Methods}

In this section, we examine how existing automatic \ac{LFQA} evaluation metrics compare to human evaluations from three aspects, namely \textit{accuracy}, \textit{robustness}, and \textit{fairness}. 
Firstly, the \textit{accuracy} is to test to what extent the automatic metrics match human judgments. Here, we use responses from seven \acp{LLM} across three benchmark datasets as testbeds to assess the alignment between automatic metrics and human evaluations.
Secondly, the \textit{robustness} is to test the reliability of the automatic metrics. We assess the stability of the outcome when subjected to minor perturbations in inputs and hyperparameters.
Finally, the \textit{fairness} is to test whether existing automatic metrics exhibit biases toward specific attributes, e.g., style, length, or topic.

%First, we analyze the \textit{accuracy} of these evaluation metrics by comparing their outputs with human evaluations. Specifically, we use responses from seven \acp{LLM} across three benchmark datasets as testbeds to assess the alignment between automatic metrics and human evaluations. 
%Next, we investigate the \textit{robustness} and \textit{fairness} of the metrics, focusing on their stability and potential biases when handling various input variations.

In the following, we first introduce the setting of the empirical study, including automatic evaluation metrics, models, and the testbed. Then, we show results concerning the \textit{accuracy}, \textit{robustness}, and \textit{fairness} of all metrics.

%In the following, we first introduce the testbed, including datasets, models, and metrics. Then, we discuss the results .

%Specifically, we firstly take responses of seven \acp{LLM} from three benchmark datasets as testbeds, and compare the evaluations of all automatic metrics with human evaluations. Then, we analyze 
% We firstly introduce the datasets used in the experiments, the models generating the answers, and the specific methods for both human and automatic evaluations. During the experiments, we first select question samples from the dataset, generate corresponding responses using different LLMs, then calculate the scores of the different responses using various evaluation metrics, and analyze the results.

\subsection{Empirical Setup}

% \subsubsection{Criteria}

% When evaluating metrics, the key question is: \textit{what constitutes a good evaluation metric?} We will address this question from three aspects: 
% \begin{itemize}
% \item Accuracy: A good metric should closely matches human judgments, reliably reflecting a model's correctness, relevance, and overall quality.
% \item Stability: A stable metric provides consistent evaluations across different prompts and input variations, remaining unaffected by minor or irrelevant changes.
% \item Fairness: A fair metric minimizes bias, treating all response types equally without favoring specific styles, lengths, or content.
% \end{itemize}

\subsubsection{Testbed}

For long-form question answering, we mainly consider two categories: one requires in-depth analysis and detailed explanations, while the other includes opinions, discussions, and other scenarios closely resembling real user interactions, which are often non-factoid. Both types of questions are considered challenging for \acp{LLM}, in terms of response generation or evaluation\cite{Krishna2021Hurdles, Fan2019eli5, zheng2024judging}. 
Specifically, we perform experiments on the following three datasets, each serving as a representative dataset for ambiguous, factoid, and open-ended question answering.
\begin{itemize}[leftmargin=*]
    \item \textit{ASQA} \cite{Stelmakh2022asqa} is an ambiguous factual questions dataset in which each question has multiple disambiguated question-answer pairs and two long-form grounded answers annotated by humans.
    \item \textit{ANTIQUE}~\cite{hashemi2020antique} is an open-ended question answering dataset, including 2626 questions asked by users of Yahoo! Answers, and relevant answers annotated by human experts. 
    \item \textit{WikiEval}~\cite{es2023ragas} is a factoid question answering dataset generated from 50 pages from Wikipedia with edits post 2022, annotated by human experts.
\end{itemize}

%\subsubsection{ASQA} \cite{Stelmakh2022asqa} is an ambiguous factual questions dataset in which each question has multiple disambiguated question-answer pairs and two long-form grounded answers annotated by humans.

%\subsubsection{ANTIQUE}\cite{hashemi2020antique} is a Non-factoid question answering dataset, including 2626 questions asked by users of Yahoo! Answers, and relevant answers annotated by human experts. 

%\subsubsection{WikiEval}\cite{es2023ragas} is a Factoid question answering dataset generated from 50 pages from Wikipedia with edits post 2022, annotated by human experts.
\noindent%
Using the datasets mentioned above, we gather responses from seven latest \acp{LLM} from five different families for analysis. The models are listed as follows:
\begin{itemize}[leftmargin=*]
    \item GLM-4-9b-chat~\cite{glm2024chatglm}: An advanced generative language model designed for dialogue applications with a focus on efficient response generation.
    \item Llama2-7b-chat~\cite{touvron2023llama2}: A conversational AI model from the Llama family, optimized for contextual understanding in dialogues with 7 billion parameters.
    \item Llama2-13b-chat~\cite{touvron2023llama2}: A more robust version of its predecessor, this model offers enhanced conversational abilities with 13 billion parameters.
    \item Llama3-8b-instruct~\cite{llama3}: Building on the Llama series, this iteration focuses on instruction-following tasks with an 8 billion-parameter architecture.
    \item GPT-3.5-instruct~\cite{OpenAI2023GPT35TurboInstruct}: A finetuned version of GPT-3.5, this model excels in following user instructions with improved precision.
    \item Mistral-7b~\cite{jiang2023mistral}: A lightweight yet powerful model, designed for streamlined AI tasks. 
%while maintaining high performance and efficiency.
    \item Solar-10.7b-instruct~\cite{kim2023solar}: A state-of-the-art model tailored for specific instruction adherence, boasting 10.7 billion parameters for diverse task performance.
\end{itemize}

\noindent%
To generate answers, we use the recommended generation parameters for each model and guide the models using few-shot settings. For each input, we select up to 8 examples. We selected 500 samples from the dev set of ASQA, 200 samples from the test set of ANTIQUE and 50 samples from WikiEval to generate answers. After filtering out invalid QA pairs-such as those resulting from model refusals-we obtained 3,500 valid samples from ASQA, 1,386 from ANTIQUE, and 350 from WikiEval. 
%The prompts and few-shot examples we used to guide the generation of answers from the seven models can be found in our repository.\footnote{\url{https://anonymous.4open.science/r/lfqa_evaluation-CE6B/src/prompts_for_answer_generation.py}} 
%For the ASQA dataset, we used the generating prompts in FLASK~\cite{Ye_2023_flask}.

\subsubsection{Meta-evaluations}
This involves the evaluation of the evaluation, serving as a method to assess the quality of automatic evaluation metrics. Here, we adopt two types of meta-evaluations, as outlined below.

\begin{itemize}[leftmargin=16pt,leftmargin=*]
\item Correlation coefficient: When comparing the differences between two distributions, a commonly used metric is the correlation coefficient. 
Following \citet{Chen_2023ref_free}, we use the Spearman correlation coefficient to measure the value correlation and Kendall correlation coefficient to measure the rank correlation between human ratings and automatic metrics.
% We follow XXX \cite{} to use the Spearman correlation coefficient and the Kendall correlation coefficient, to measure the relationship between human ratings and automatic metrics.
\item Win Rate and Agreement: When comparing multiple LLMs, researchers often use the win rate, defined as the fraction of instances in which model A's response outperforms model B's.
Furthermore, for two different evaluators, their agreement can be calculated, which represents the proportion of instances where the evaluators made the same judgment regarding the quality of the responses from the two models~\cite{zheng2024judging}.
% \item Robustness Rate (RR): RR measures the consistency of the LLM's judgments before and after introducing bias into the candidate answers. For each sample in the test dataset ($D$), the LLM judge performs two evaluations: one without perturbation to obtain ($y_i$), and another with a biased perturbation to obtain ($\hat{y}_i$). RR is then calculated as the proportion of instances where the LLM's judgment remains consistent, i.e., ($y_i = \hat{y}_i$), across all samples in ($D$). A higher RR indicates greater robustness, meaning the LLM's judgments are less affected by the introduced bias, while a lower RR suggests susceptibility to bias\cite{JusticeorPrejudice_2024}. 
\end{itemize}

\begin{figure}
    \centering
    \includegraphics[width=0.95\columnwidth]{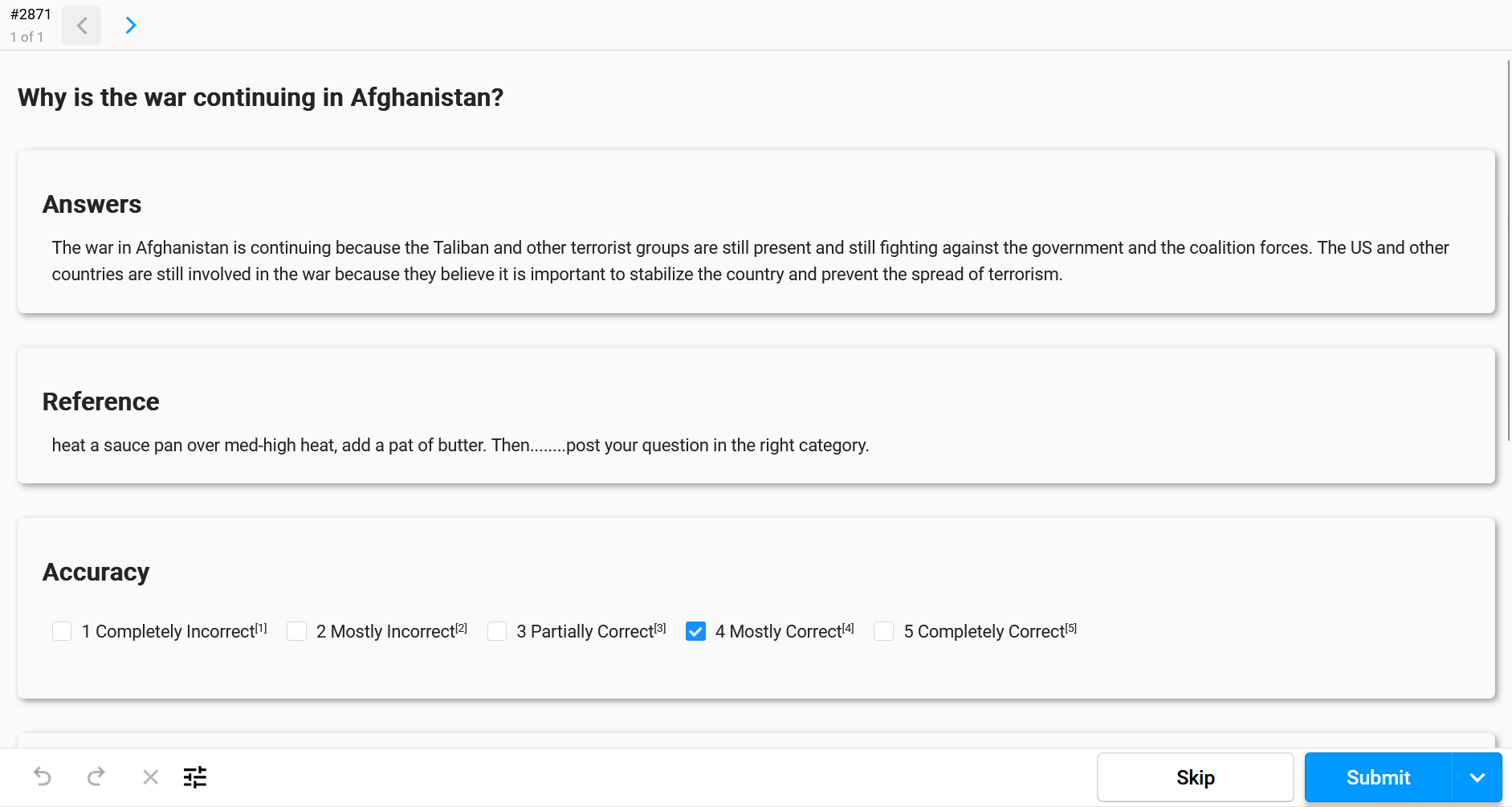}
    \caption{Interface used for collecting human annotations.}
    \label{fig:fig_interface}
\end{figure}

\input{tables/table_acurracy}

\subsubsection{Human annotations}
\label{sec:human_annotations}
In the annotation of model responses, we drew inspiration from previous work~\cite{Zhang_2023_llmeval} and primarily considered two evaluation aspects.
\begin{itemize}[leftmargin=*]
\item Correctness aims to determine the accuracy of the answers, determining whether there are any factual errors or inaccuracies in the model's response. A high-quality response should be factually reliable and free from mistakes.
\item Informativeness examines whether the response contains sufficient and relevant information, identifying any missing or omitted details, and ensuring there is no excessive redundancy. A high-quality response should should be informative enough without being redundant.
\end{itemize}

\noindent%
For the results of the seven models on the ASQA, ANTIQUE, and WikiEval datasets, we randomly sampled 50, 200, and 50 questions, respectively, for manual evaluation, with each question having 7 answers from different models. After removing a few cases where models refused to answer or generated incorrect responses, we collected 343, 1386, and 350 valid QA pairs that include both correctness and informativeness. We developed a web-based interface to streamline the annotation process. Figure~\ref{fig:fig_interface} illustrates the interface for a single question. Each question is paired with a generated answer and a reference answer. 

% Our annotation interface is shown in the appendix \ref{app:human_eval_interface}.

\subsubsection{Automatic metrics}
For evaluation metrics, we include $7$ widely used metrics for long-form question answering, including traditional deterministic metrics and recently model-based metrics.

\textbf{Deterministic metrics}: The deterministic metrics typically assess results based on lexical overlap between the predicted answer and the reference answer.
Early works on \ac{LFQA} usually take \textit{Exact Match} (EM) and \textit{Rouge-L} (RL) as the evaluation metrics \cite{Stelmakh2022asqa, Fan2019eli5}. The EM metric is to evaluate whether the predicted answer matches any of a fraction in the reference answer \cite{Stelmakh2022asqa}.
The Rouge-L is to evaluate the quality of the predicted answer by measuring the longest common subsequence between the generated text and the reference answer \cite{lin2004rouge}.
We also include the \textit{Disambig-F1} (DF1) \cite{Stelmakh2022asqa}, which is to evaluate the fraction of diambiguated questions answered by the predicted answer, for the evaluation study. 
%which is also widely used in the ASQA dataset, for
%For ASQA dataset, we follow the original settings of the dataset, using Exact Match (EM), Rouge-L, and Disambig-F1 to measure the correctness of the answers. 

\textbf{Model-based metrics}: The model-based metrics often assess the semantic accuracy of the generated answer using pre-trained models, extending beyond traditional n-gram matching.
The \textit{BERTScore} (BS) \cite{zhang2019bertscore} calculate the cosine similarity between the generated answer and the reference answer.
In addition, we considered the answer relevance (AR) from RAGAS~\cite{es2023ragas} to supplement the measurement of responses' informativeness. 
For \ac{LLM}-based evaluation, we follow LLM-EVAL~\cite{lin2023llm_eval} to use prompt-driven GPT-4 model as the judge in both coarse-grained (CG) and fine-grained (FG) settings.

%To compare the performance of embedding-based metrics, we additionally include BERTScore in our evaluation. For ANTIQUE dataset, we evaluate all answers using Rouge-L and BERTScore. Since the questions in the ANTIQUE dataset do not have explicit reference answers, we select the original best answers collected from community websites as references. The majority of these are labeled as having the highest relevance (Label 4). It is important to note that these "best answers" may not accurately address the questions, which we will discuss in detail in the Section~\ref{sec:results}.

% \textbf{Model-based metrics}: For the model responses on the two datasets, we referred to the LLM-EVAL and used a prompt-driven GPT-4 model to evaluate them in coarse-grained and fine-grained settings. The prompt templates and different prompt instructions are shown in Figure \ref{fig_prompt_template}, Figure \ref{fig_cg}, Figure \ref{fig_fg}. In addition, we considered the answer relevance method from RAGAS\cite{es2023ragas} to supplement the measurement of responses' informativeness.

%\section{Results}\label{sec:results}

%In this section, we compare different evaluation methods, analyzing their consistency and biases.

\subsection{Accuracy of Automatic Metrics}
% Accuracy
% RQ1:
% to provide an answer to RQ1:
In this section, we try to answer RQ1, that is to what extent do existing automatic evaluation metrics serve as substitutes for human evaluations in \ac{LFQA}. In this part, we take responses of seven \acp{LLM} on three datasets as the testbed. For each response, we conduct a human rating and compute the scores for each metric. We then assess the correlation coefficient between these metrics and human ratings across the entire dataset. Additionally, for any pair of model-generated responses, we calculate the agreement between each metric and human rating to determine whether they align in their preference between these two answers. The results of correlation coefficients between automatic metrics and human ratings for the three datasets are presented in Table~\ref{tab:asqa_main} and~\ref{tab:antique_main}. 
%We compared the results of these metrics across responses from different models and observed several notable trends. 

Firstly, for deterministic metrics, we can see that exact match demonstrates relatively strong alignment with human evaluation on ASQA dataset, even outperforming LLM-based evaluations such as GPT-4o, Claude-3.5 and Gemini-2.0. This is primarily because exact match focuses solely on assessing short answers in ASQA. However, while evaluating long answers, metrics like Rouge-L and Disambig-F1 exhibit poor consistency with human evaluations. 
An exception is observed with Rouge-L, which yields significantly different results on the Antique and WikiEval datasets. It shows high consistency with human evaluations on WikiEval but low consistency on Antique. This discrepency arises because Antique features non-factoid \ac{QA} with open-ended answers, whereas WikiEval focuses on factoid \ac{QA} with closed-ended answers.

Secondly, for model-based metrics, we can see that BertScore exhibits a trend similar to Rouge-L, showing low consistency with human evaluations on the ASQA and Antique datasets but high consistency on the WikiEval dataset.
In contrast, LLM-based evaluations exhibit strong consistency with human evaluations across various styles of \ac{LFQA}, including the ambiguous \ac{QA}, fatoid \ac{QA}, and non-factoid \ac{QA}, highlighting their stability in evaluation. For example, the fine-grained (FG) GPT4 achieves the best performance on both the ASQA and Antique datasets. 

Lastly, when comparing different LLM-based evaluations, we find that GPT-4o evaluation demonstrates superior performance over other \acp{LLM}. For example, the GPT-4o, Claude-3.5, and Gemini-2.0 achieve spearman correlation score of $42.0$, $33.0$, and $32.4$, respectively, when compared to human evaluations. Moreover, the fine-grained evaluation with GPT-4o improves its score from $42.0$ to $55.0$, which demonstrates the importance of providing more detailed instructions for LLM-based evaluations.

In summary, LLM-based evaluations demonstrate stability across different types of \ac{LFQA}, whereas deterministic evaluations are less reliable for open-ended \ac{QA}. Furthermore, conducting fine-grained evaluations with LLM-based assessments would produce more nuanced results.

\input{tables/table_temperatures}

\subsection{Robustness of Automatic Metrics}

In this section, we focus on the first part of RQ2, which investigates the robustness of automatic evaluation metrics. 
For this purpose, we introduce small perturbations to each evaluation method, and analyze the changes before and after the perturbation. Here, we concentrate on model-based evaluation metrics.
Firstly, we collects the scores generated by automatic evaluation metrics under their default configurations. Subsequently, we introduce perturbations to the experimental conditions and obtain the corresponding perturbed scores. To quantify the consistency of evaluators' decisions before and after the introduction of perturbations, we statistically analyzed the model win rates and the distribution of metric scores under different experimental settings.

% Stability
% 1. perturbation of prompts
% 2. perturbation of hyper-parameters

\begin{figure}[!t]
    \centering
    \includegraphics[width=0.8\columnwidth]{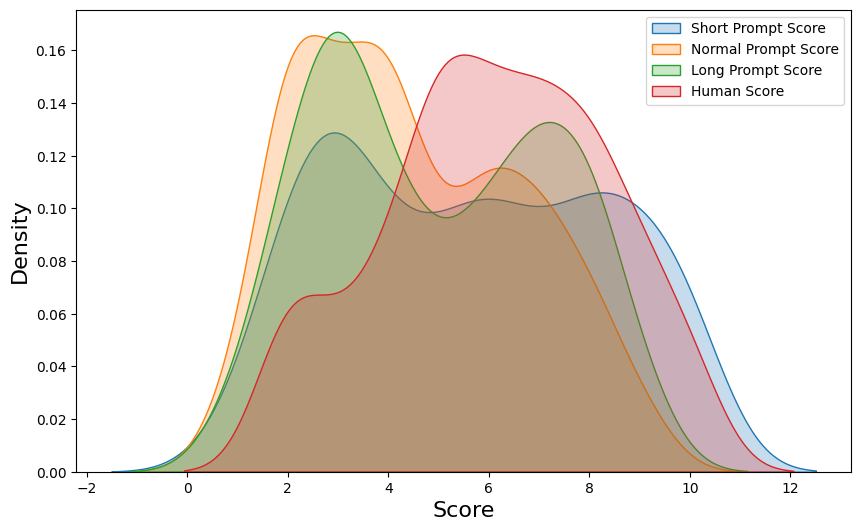}
    \caption{Score distribution of different prompts on ASQA.}
    \label{fig:asqa_kde_prompt}
\end{figure}

\subsubsection{Perturbation of prompts}
\label{sec:Perturbation_of_prompts}
The prompts of \acp{LLM} have been shown to significantly influence their output, giving rise to a new research area known as prompt engineering \cite{ye2023prompt}. To better understand the impact of prompts on the evaluation capability of \acp{LLM}, we conducted experiments comparing different prompts using GPT-4o on the ASQA dataset, including short prompt, normal prompt, and long prompt. More detailed description of prompts are shown in code repository.\footnote{\url{https://github.com/bugtig6351/lfqa_evaluation/src/prompts.py}}

The results are depicted in Figure \ref{fig:asqa_kde_prompt}:
\begin{enumerate*}[label=(\roman*)]
\item LLM-based evaluations tend to yield lower scores compared to human evaluations, with the majority of LLM-based scores clustering around $3$, while human evaluation scores are predominantly around $5$. This discrepancy may be due to LLM-based evaluations being more stringent than human evaluations. 
\item The length of the prompt has a significant impact on the evaluation scores of \acp{LLM}. It is evident that short prompts tend to result in more high scores. For example, the number of samples with scores above 4 for the short prompt is noticeably higher than for the normal and long prompts.
\end{enumerate*}

\input{figures/fairness_length}

\subsubsection{Perturbation of hyper-parameters}

Hyperparameters such as sampling temperature, top-k sampling, repetition penalty, and maximum token length all play a role in shaping the LLM's output and overall performance \cite{OpenAI_Platform}. 
% Generally, temperature is the most frequently used option among these, as it controls the randomness during the inference process\cite{OpenAI_Platform, Claude_Messages, Gemini_API}. 
However, the impact of sampling temperature on LLM-as-a-judge has not been specifically investigated. The selection of sampling temperature is largely based on guesswork and intuition. 
%For example, lower temperatures are believed to favor more deterministic, less open-ended, or less creative prompts, while higher temperatures tend to produce more creative outcomes.
Here, we take GPT-4o as the judge to analyze the impact of temperature.

All results are summarized in Table~\ref{temperature_analysis}. If multiple models have the same score, we take the average rank as their shared rank. As we can see:
\begin{enumerate*}[label=(\roman*)]
\item In the ASQA dataset, the evaluation results are highly stable, where minimal changes in score and no changes in the rankings of any \acp{LLM}. This could be because the evaluation score for each model vary widely, ranging from $3.03$ to $8.66$, making it difficult for the rankings to change.
\item In the WikiEval dataset, the evaluation results vary significantly with changes in the temperature of the \ac{LLM} judge. For example, the rank of GPT-4o drops from 4th to 6th when the temperature increases from $0.0$ to $0.3$. In contrast, its rank improves from 4th to 2nd when the temperature is set to $0.7$ or $1.0$. This may be because the evaluation scores of each model are very similar, with the minimum and maximum values being $6.60$ and $8.62$, respectively.
\end{enumerate*}
Overall, the above observations indicate that the temperature of the \ac{LLM} does impact the evaluation, but the extent of this impact depends on the dataset.
% changed a lot as we change the temprature of the LLM judge. For example, the rank of the gpt-4o model drops from rank 4 to rank 6 when the temperature is changed from 0.0 to 0.3, and it improves from rank 4 to rank 2 when the temperature is changed from 0.0 to 0.7 or 1.0.
% the closed-source \acp{LLM} performs much better than open-sourced \acp{LLM}, where cloude-3.5  2) ... 3) ...

% consistency between different temperatures
% influence on model winrates

% \begin{figure}
%     \centering
%     \includegraphics[width=\columnwidth]{figures/length_box_wikieval.png}
%     \caption{Relationship between Answer Length and Different Metrics on Wikieval}
%     \label{fig:length_box_wikieval}
% \end{figure}

% \begin{figure}
%     \centering
%     \includegraphics[width=\columnwidth]{figures/length_box_asqa.png}
%     \caption{Relationship between Answer Length and Different Metrics on ASQA}
%     \label{fig:length_box_asqa}
% \end{figure}

% \begin{figure}
%     \centering
%     \includegraphics[width=\columnwidth]{figures/length_box_antique.png}
%     \caption{Relationship between Answer Length and Different Metrics on ANTIQUE}
%     \label{fig:length_box_antique}
% \end{figure}

\subsection{Fairness of Automatic Metrics}
% Fairness
% 1. length bias
% 2. question category
% 3. self-reinforcing
% 4. word expression

In this section, we address the second part of RQ2, which focuses on the fairness of automatic evaluation metrics. Specifically, we conduct a comprehensive analysis of the evaluators' biases across four key dimensions: response length, question type, answer generation models, and language representations.

\subsubsection{Length bias} 

\begin{table}[!t]
\centering
\caption{Question types in the ANTIQUE dataset.}
\label{tab:type}
\begin{tabular}{lrr}
\toprule
{Question Type} & {Count} & {Proportion} \\
\midrule
REASON & 427 & 0.31 \\
INSTRUCTION & 357 & 0.26 \\
EVIDENCE-BASED & 266 & 0.19 \\
EXPERIENCE & 154 & 0.11 \\
DEBATE & 105 & 0.08 \\
COMPARISON & 77 & 0.06 \\
\bottomrule
\end{tabular}
\end{table}

Here, we aim to examine whether answer length influences the evaluation of automatic metrics, particularly LLM-based evaluation methods. 
To this end, we divide all answers into five bins with equal sample sizes based on their length and analyze the performance of different evaluation metrics across these length intervals.
All the results are depicted in Figure~\ref{fig:length_box}.
Our experimental results reveal distinct patterns in the relationship between answer length and various evaluation metrics. 

\begin{figure}
    \centering
    \includegraphics[width=\columnwidth]{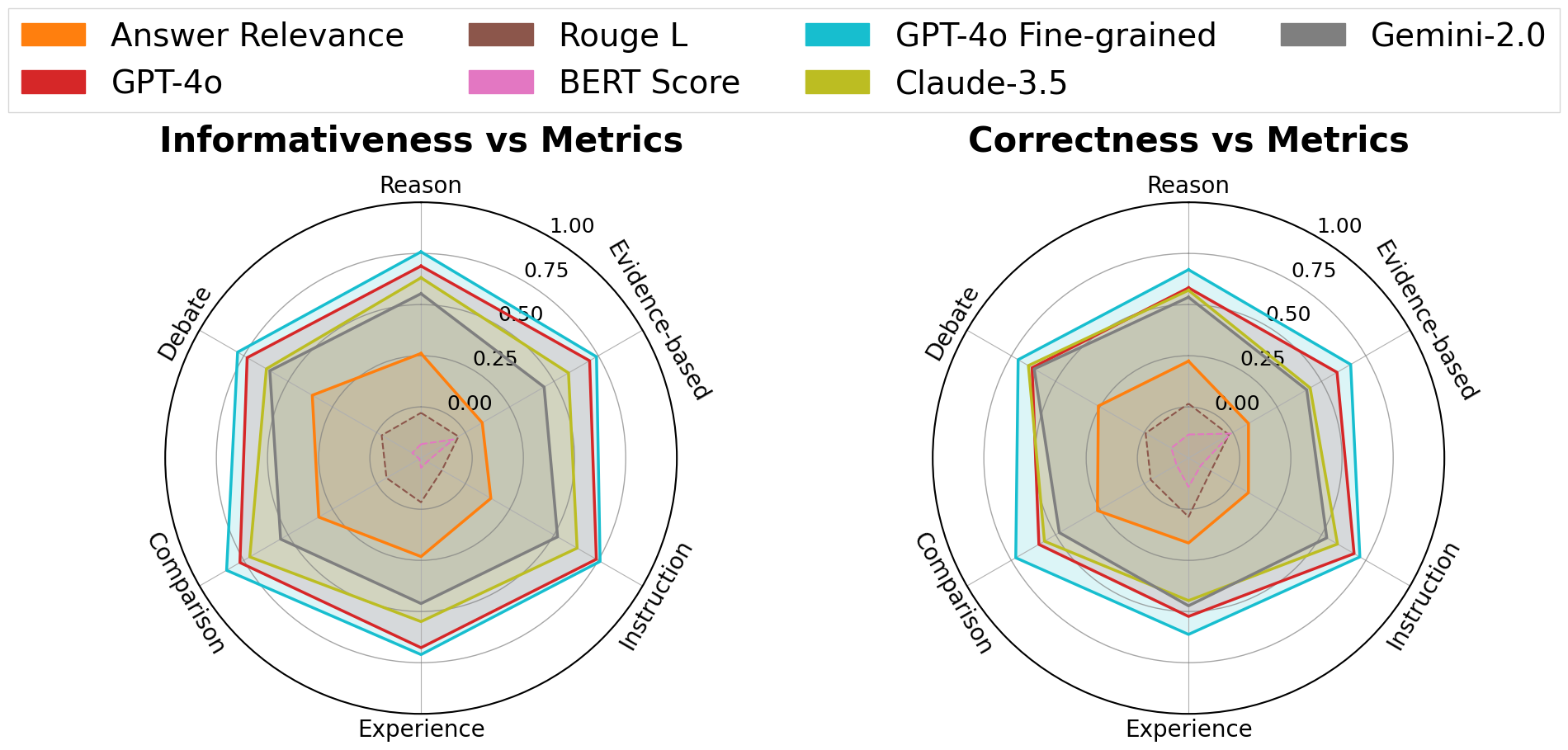}
    \caption{Relationship between different metrics and human evaluations across different question types on the ANTIQUE dataset (left: Informativeness, right: Correctness).}
    \label{fig:type}
\end{figure}

% \begin{figure*}
%     \centering 
%     \includegraphics[width=\textwidth]{figures/avg_idf_wikieval.png}
%     \caption{Relationship between Answer IDF and different metrics on the WikiEval dataset.}
%     \label{fig:answer_idf}
% \end{figure*}
\begin{figure*}
    \centering 
    \begin{subfigure}{\textwidth}
    \includegraphics[width=\textwidth]{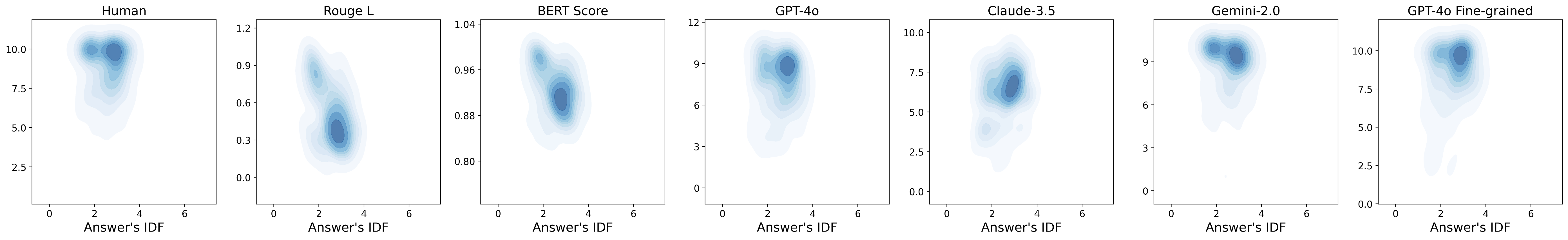}
    \caption{Results on the WikiEval dataset.}
    \label{fig:avg_idf_wikieval}
    \end{subfigure}

    %\vspace{0.1cm}

    \begin{subfigure}{\textwidth}
    \includegraphics[width=\textwidth]{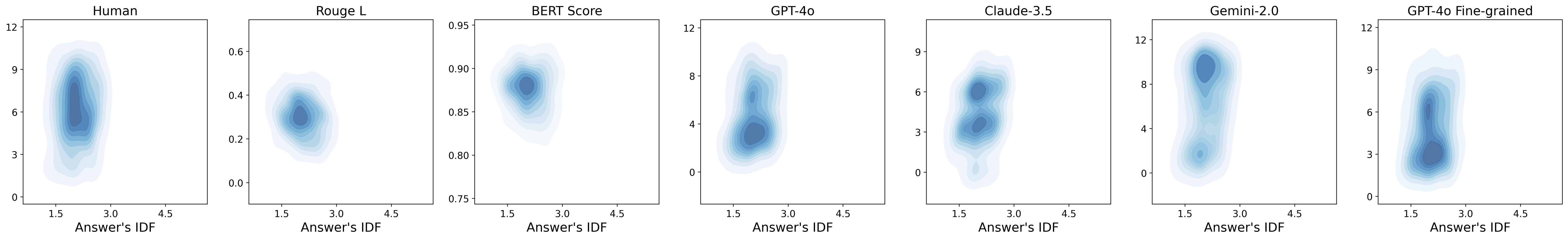}
    \caption{Results on the ASQA dataset.}
    \label{fig:avg_idf_asqa}
    \end{subfigure}
    
    %\vspace{0.1cm}

    \begin{subfigure}{\textwidth}
    \includegraphics[width=\textwidth]{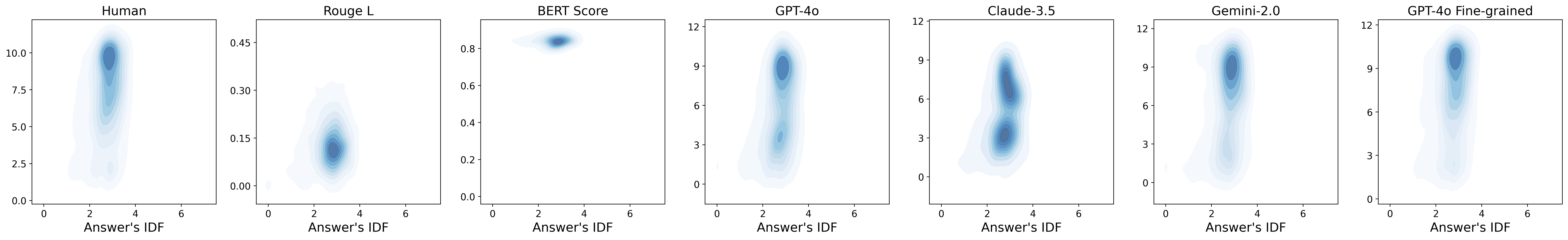}
    \caption{Results on the Antique dataset.}
    \label{fig:avg_idf_antique}
    \end{subfigure}

    \caption{Relationship between the IDF of answer and different metrics.}
    \label{fig:answer_idf}
\end{figure*}

We observe a negative correlation between answer length and scores on the Rouge-L and BERTScore metrics. As the length of the answer increases, the Rouge-L and BERTScore metrics tend to decrease. This trend suggests that longer answers may introduce redundancy or deviate from the concise, information-dense content that these metrics favor. Rouge-L, which measures the overlap between generated and reference texts, is particularly sensitive to extraneous information that dilutes the precision of the answer. Similarly, BERTScore, which evaluates semantic similarity, may penalize longer answers that include tangential or less relevant content. 

On the ASQA and WikiEval dataset, we observe a positive correlation between answer scores and length in Claude-3.5. While the average scores remained relatively stable, the lower bounds of scores assigned by GPT-4 and Gemini-2.0 increased significantly with longer answers. For the Antique dataset, the trend of increasing scores with length is evident, and this trend is more pronounced in LLM-based methods compared to human evaluations. This indicates that LLMs are less likely to assign low scores to longer answers, a phenomenon not observed in human evaluations. These evaluation frameworks may prioritize comprehensiveness and detail over conciseness.

Our analysis reveals that answer length significantly influences the performance of automatic evaluation metrics, with distinct patterns observed across different metrics and models. Combine traditional metrics (e.g., Rouge-L, BERTScore) with LLM-based metrics to balance the strengths and weaknesses of each approach. To address these biases and improve the robustness of automatic evaluation metrics, some strategies could be considered. For example, traditional metrics can ensure precision and conciseness, while LLM-based metrics can capture comprehensiveness and detail. And develop metrics that account for answer length by normalizing scores based on the amount of relevant information. This could involve penalizing redundancy while rewarding additional meaningful content.

% \begin{figure*}
%     \centering
%     \includegraphics[width=\textwidth]{figures/length_box_asqa.png}
%     \caption{Relationship between Answer Length and Different Metrics on ASQA}
%     \label{fig:length_box_asqa}
% \end{figure*}

% \begin{figure}
%     \centering
%     \includegraphics[width=\columnwidth]{figures/asqa_lines.png}
%     \caption{Average win rate of seven models under different metrics on ASQA(\textit{GPT4 CG} for coarse-grained GPT-4 scoring, \textit{GPT4 FG} for fine-grained GPT-4 scoring, \textit{DR Score} for DisambigF1-RougeL Score}
%     \label{fig:asqa_lines}
% \end{figure}

% \begin{figure}
%     \centering
%     \includegraphics[width=\columnwidth]{figures/antique_lines.png}
%     \caption{Average win rate of seven models under different metrics on ANTIQUE(\textit{GPT4 CG} for coarse-grained GPT-4 scoring, \textit{GPT4 FG} for fine-grained GPT-4 scoring,\textit{GPT4 FGC} for fine-grained GPT-4 scoring with Category informations)}
%     \label{fig:antique_lines}
% \end{figure}

\input{figures/self-reinforcing-heatmap}

\subsubsection{Question type}

In this section, we investigate the potential biases of different evaluation metrics when assessing various types of questions. Building upon the research conducted by \citet{Bolotova2022taxonomy}, we classified 200 questions from the antique dataset, with the specific quantities and proportions detailed in Table~\ref{tab:type}. For each category of questions, we calculated the kendall correlation consistency between human evaluators' scores for correctness and informativeness and various metrics, as illustrated in Figure~\ref{fig:type}.

Our analysis reveals several key findings. First, we observe that LLM-based metrics significantly outperform deterministic metrics (Rouge-L and BERTScore), with the latter showing negative correlation coefficients on this dataset, rendering them practically unreliable for evaluation purposes. Second, LLM-based metrics demonstrate slightly higher consistency in assessing informativeness compared to correctness, particularly for comparison and experience-type questions. This discrepancy may stem from the inherent difficulty of LLMs in identifying factual inaccuracies within generated responses. Finally, evidence-based and experience-type questions emerge as challenging areas for all metrics, with the three primary evaluation models (GPT-4, Claude-3.5, and Gemini-2.0) consistently underperforming relative to their average performance across other question types.

Based on our findings, we can conclude that LLM indeed exhibit evaluation biases across different types of question, particularly for questions relying on precise factual information or personal experiences and recommendations. However, potential improvements can be achieved through the implementation of more granular prompting strategies combined with a diversified evaluation approach.

% \begin{figure}
%     \centering 
%     \begin{subfigure}{\columnwidth}
%     \includegraphics[width=0.9\columnwidth]{figures/avg_idf_wikieval.png}
%     \caption{Results on the WikiEval dataset.}
%     \label{fig:avg_idf_wikieval}
%     \end{subfigure}

%     \vspace{0.2cm}

%     \begin{subfigure}{\columnwidth}
%     \includegraphics[width=\columnwidth]{figures/avg_idf_asqa.png}
%     \caption{Results on the ASQA dataset.}
%     \label{fig:avg_idf_asqa}
%     \end{subfigure}
    
%     \vspace{0.2cm}

%     \begin{subfigure}{\columnwidth}
%     \includegraphics[width=\columnwidth]{figures/avg_idf_antique.png}
%     \caption{Results on the Antique dataset.}
%     \label{fig:avg_idf_antique}
%     \end{subfigure}

%     \caption{Relationship between Answer IDF and different metrics.}
%     \label{fig:answer_idf}
% \end{figure}

\input{tables/table_enhance}
\input{tables/table_prompt_settings}

\subsubsection{Self-reinforcing}
In this section, we investigate whether LLM-based metrics show a preference for results generated by themselves. We take the widely adopted \acp{LLM}, i.e., GPT-4o, Claude-3.5, and Gemini-2.0, as both evaluators and generators on the ASQA dataset. Specifically, we take these three \ac{LLM} to generate answers for each question, and conduct a pairwise comparison between them and other seven open-sourced \acp{LLM}. The results are illustrated in Figure~\ref{fig:self_reinforce}.

We can see that: 
\begin{enumerate*}[label=(\roman*)]
\item The three closed-source \acp{LLM} (GPT-4o, Claude-3.5, and Gemini-2.0) consistently outperform the seven baseline \acp{LLM} (Mistral-7b, Llama2-7b, Llama3-8b, Solar-10.7b, GLM4-9b, and GPT-3.5-turbo). Among them, Claude-3.5 achieves the best performance in all baselines across all three LLM-based evaluations. 
\item Among the three evaluation methods, we observe that GPT-4o and Claude-3.5 evaluations exhibit a strong bias towards their own responses. For example, GPT-4o achieves significantly higher win rates against all baselines when evaluated using GPT-4o compared to evaluations conducted by Claude-3.5 and Gemini-2.0. Similarly, Claude-3.5 demonstrates much higher win rates in its own evaluation than when assessed by GPT-4o and Gemini-2.0. Furthermore, the Gemini-2.0 evaluation also gives much higher scores to itself compared to GPT-4o and Claude-3.5 evaluations (e.g., $0.468$ vs.\ $0.262$/$0.256$, $0.388$ vs.\ $0.163$/$0.06$), when compared with GPT-4o and Claude-3.5.
\end{enumerate*}

In summary, LLM-based evaluations tend to assign significantly higher scores to their own outputs, demonstrating a clear evaluation bias.
Interestingly, despite this bias, the models' ranking remains consistent across different evaluations. Based on these, we recommend the following: 
\begin{enumerate*}[label=(\roman*)]
\item Prioritize using rankings over scores for comparisons, 
\item Employ multiple evaluation methods to achieve more reliable and stable results.  
\end{enumerate*}

% \begin{figure}
%     \centering
%     \includegraphics[width=\columnwidth]{figures/asqa_self_reinforcing_heatmap_gpt_4o.png}
%     \caption{Win rate of 10 models under different metrics by GPT-4o on ASQA}
%     \label{fig:asqa_self_reinforcing_heatmap_gpt_4o}
% \end{figure}

% \begin{figure}
%     \centering
%     \includegraphics[width=\columnwidth]{figures/asqa_self_reinforcing_heatmap_claude.png}
%     \caption{Win rate of 10 models under different metrics by Claude-3.5 on ASQA}
%     \label{fig:asqa_self_reinforcing_heatmap_claude}
% \end{figure}

% \begin{figure}
%     \centering
%     \includegraphics[width=\columnwidth]{figures/asqa_self_reinforcing_heatmap_gemini.png}
%     \caption{Win rate of 10 models under different metrics by Gemini-2.0 on ASQA}
%     \label{fig:asqa_self_reinforcing_heatmap_gemini}
% \end{figure}

\subsubsection{Word expression}

Prior work shows that \acp{LLM} often produce verbose, formal content \cite{wu2025survey}.
To investigate whether LLM-based evaluations favor answers containing rarer terms, we compute each answer's  average \ac{IDF} and examine its correlation with the evaluation scores. 
The results are illustrated in Figure \ref{fig:answer_idf}.
%The results for the WikiEval dataset are presented in Figure \ref{fig:avg_idf_antique}. Similar patterns were observed on the ASQA and Antique datasets, with detailed results available at \url{https://github.com/bugtig6351/lfqa_evaluation/}. As observed: Human scores are evenly distributed across \ac{IDF} values, showing no clear link between the quality of the answer and the rarity of the terms. Automatic metrics like Rouge-L and BERTScore strongly correlate with IDF values, following near-normal distributions. In contrast, GPT-4o and Gemini-2.0 align more closely with human judgments. Notably, under fine-grained settings, GPT-4o tends to favor responses with higher \ac{IDF} values, indicating a potential bias toward lexically sophisticated responses. These results highlight the importance of dataset-specific evaluation strategies and the need to address IDF-related biases in LLM-based metrics, and this similarity becomes even more pronounced when fine-grained evaluation settings are applied.
As observed: On Antique and WikiEval, human scores are evenly distributed across \ac{IDF} values, showing no clear link between answer quality and term rarity, while ASQA displays a unique bimodal pattern, suggesting different answer characteristics. Automatic metrics like Rouge-L and BERTScore strongly correlate with IDF values, following near-normal distributions on Antique and WikiEval. In contrast, GPT-4o and Gemini-2.0 align more closely with human judgments. Notably, under fine-grained settings, GPT-4o tends to favor responses with higher \ac{IDF} values, indicating a potential bias toward lexically sophisticated responses. These results highlight the importance of dataset-specific evaluation strategies and the need to address IDF-related biases in LLM-based metrics, and this similarity becomes even more pronounced when fine-grained evaluation settings are applied.

%the human evaluation do not show any particular preferences on answers with  
% To analyze the relationship between lexical expressions and different metrics, we calculated the inverse document frequency (IDF) across all datasets， and examined the relationship between answers' average IDF and the scores of various metrics, as illustrated in Figure~\ref{fig:avg_idf_antique}. Similar results were also observed on the other two datasets, and further details can be found in Appendix~\ref{app:word_expression}.

\section{Method}
\label{sec:method}
% RQ3
% improving metrics
\input{tables/table_instruction}

In this section, we investigate approaches to improve the performance of LLM-based evaluators to address RQ3: ``How can the effectiveness and robustness of existing evaluation methods be improved?''
% through systematic experimentation with various prompting strategies. 
To achieve this, we transfer the findings in previous sections into criteria in the prompt of LLM-based evaluation.
Specifically, we decompose the prompt architecture into four key components: task description, data specifications, output requirements, and evaluation criteria. Then, we obtain 9 prompts by different combinations of these four components as outlined in Table~\ref{tab:prompts}. Due to space limitations, all detailed prompts can be found in our released code.\footnote{\url{https://github.com/bugtig6351/lfqa_evaluation/src/prompts.py}}% The instruction adopted is as shown in Table~\ref{tab:instructions}.
% As outlined in Table~\ref{tab:prompts}, 
Finally, we measure performance variations by calculating the Kendall correlation coefficient changes between GPT-4o evaluations and human ratings for each prompt configuration.

All results are shown in Table~\ref{tab:enhance}. Our main findings are as follows.
\begin{enumerate}[leftmargin=*,label=(\arabic*)]
    \item \textbf{All four components are essential for improving LLM-based evaluations}: Comparing the performance of P1 with P2, P5 and P6, it can be observed that P1 achieves significantly better performance across all \acp{LLM}. P1 incorporates all four components, while P2 excludes the \textit{Criteria}, P5 omits the \textit{Data}, and P6 lacks the \textit{Task}. Moreover, the evaluation of glm4-9b and gpt-turbo-3.5 drop significantly when removes the \textit{Criteria} from the prompts (i.e., P2 vs.~P3). Besides, the \textit{Task} is important for LLM-based evaluations as performance drops significantly by comparing P3 vs.~P6.
    
    \item \textbf{The order of components influences the LLM-based evaluations}: A comparison of P1, P3, P4, P7, and P9 shows that the order of the four components has varying impacts on the evaluation of different \acp{LLM}. Notably, the placement of the \textit{Output} appears to have minimal influence on gpt-3.5-turbo, as the performance is very close between P3 and P4—where \textit{Output} is positioned last in P3 and first in P4.
    
    \item \textbf{No consistent improvements can be obtained for the evaluation of all \acp{LLM} by one strategy}: There is no one strategy can boost the performance of all \acp{LLM} evaluations, where glm4-9b, llama2-7b, and llama3-8b performs best with P4, gpt-3.5-turbo performs best with P3, llama2-13b performs best with P1, solar-10.7b performs best with P9. The differences in answer styles between models may significantly affect the evaluator's performance across various prompts.
\end{enumerate}
These results demonstrate the necessity of structured guidance and precise evaluation guidelines for reliable assessments.

\section{Conclusion}

In this work, we evaluated a range of automatic evaluation metrics, including traditional deterministic metrics and model-based metrics, across three diverse datasets: ASQA, Antique, and WikiEval. For each dataset, we compared the scores generated by these metrics with human evaluations, analyzed their robustness under perturbations, and investigated potential biases related to answer length, question type, self-reinforcement, and language expression. Additionally, we explored methods to improve the performance of LLM-based evaluators through fine-grained evaluation strategies.

Our evaluation indicates that even with relatively high-quality reference answers, deterministic evaluation metrics still perform poorly and often do not exhibit significant correlation with human judgments. Their performance is highly dependent on the dataset and question type, with limited applicability to complex LFQA tasks. LLM-based metrics, such as GPT-4o and Claude-3.5, demonstrate better alignment with human evaluations and greater stability across different types of questions. However, they are susceptible to biases related to answer length, question type, and self-reinforcement. For example, LLM-based evaluators tend to favor longer answers and their own generated responses.
The observed biases in LLM-based evaluations call for the development of more equitable evaluation frameworks that account for factors such as answer length, question type, and language expression.

% \mdr{What should we do next?} More fine-grained evaluations can somewhat mitigate this issue, but further investigation is needed.

%\section{Limitations and Future direction}
Although our study has investigated the performance of automatic evaluation metrics across multiple dimensions, there are still some limitations that point to directions for future work: (i) Our 750-question evaluation covers only 3 QA categories, while real-world LFQA involves more diverse formats (e.g., multihop reasoning). Future research should evaluate automatic metrics on broader datasets, including more diverse type and domain of questions. (ii) We adopted a single-point scoring approach, and for human evaluation, we only set two main scoring criteria: correctness and informativeness. This may miss subtle quality differences between answers. 
A more nuanced evaluation setup would help fully capture various potential responses and edge cases. (iii) Our experiments have revealed the sensitivity of LLM-based models to prompts, but we have not conducted a detailed analysis of how different models respond to this, further investigation is needed.

\begin{acks}
This work was funded by the National Natural Science Foundation of China (NSFC) under Grants No. 62372431, 62472408 and 62441229, the Strategic Priority Research Program of the CAS under Grants No. XDB0680102 and XDB0680301, the National Key Research and Development Program of China under Grants No. 2023YFA1011602,
%the Youth Innovation Promotion Association CAS under Grants No. 2021100, 
the Lenovo-CAS Joint Lab Youth Scientist Project, and the project under Grants No. JCKY2022130C039.
This research was (partially) supported by the Dutch Research Council (NWO), under project numbers 024.004.022, NWA.1389.20.\-183, and KICH3.LTP.20.006, and the European Union's Horizon Europe program under grant agreement No 101070212.
All content represents the opinion of the authors, which is not necessarily shared or endorsed by their respective employers and/or sponsors.
\end{acks}

\bibliographystyle{ACM-Reference-Format}
\bibliography{main}

\end{document}

%% file: preamble/packages.tex
\usepackage{multirow}
\usepackage{adjustbox}
\usepackage{pgfplots}
\usepgfplotslibrary{colormaps} 
\usepackage{pgfplotstable}
\usepackage{colortbl}
\usepackage{array}
\usepackage{booktabs}
\usepgfplotslibrary{groupplots}
\usetikzlibrary{patterns}
\usepackage{acronym}
\usepackage[inline]{enumitem}
\usepackage{colortbl}
\usepackage{booktabs}
\usepackage{xcolor}
\usepackage{graphicx}
\usepackage{makecell}
\usepackage{subcaption}
\usepackage[skip=2pt]{caption}
\usepackage{tcolorbox}
\usepackage{tabularx}

%% file: preamble/definitions.tex
\AtBeginDocument{%
  }

\makeatletter
\newcommand\headingnodot{\def\@toclevel{4}%
  \@startsection{paragraph}{4}{\z@}%
  {-.2\baselineskip \@plus -2\p@ \@minus -.2\p@}%
  {-3.5\p@}%
  {\ACM@NRadjust{\bfseries}}}
\newcommand{\heading}[1]{\headingnodot{#1.}}
\makeatother

\acrodef{QA}{question answering}
\acrodef{LLM}{large language model}
\acrodef{LFQA}{long-form question answering}
\acrodef{IDF}{inverse document frequency}

%% file: preamble/authors.tex
\author{Ning Xian}
\authornote{Both authors contributed equally to this research.} 
\orcid{0009-0004-1220-3021}
\affiliation{%
  \institution{Institute of Computing Technology, Chinese Academy of Sciences}
  \city{Beijing}
  \country{China}
}
\email{xianning21s@ict.ac.cn}

\author{Yixing Fan}
\authornotemark[1]
\orcid{0000-0003-4317-2702}
\affiliation{%
  \institution{Institute of Computing Technology, Chinese Academy of Sciences}
  \city{Beijing}
  \country{China}
}
\email{fanyixing@ict.ac.cn}

\author{Ruqing Zhang}
\orcid{0000-0003-4294-2541}
\affiliation{%
  \institution{Institute of Computing Technology, Chinese Academy of Sciences}
  \city{Beijing}
  \country{China}
}
\email{zhangruqing@ict.ac.cn}

\author{Maarten de Rijke}
\orcid{0000-0002-1086-0202}
\affiliation{%
  \institution{University of Amsterdam}
  \city{Amsterdam}
  \country{The Netherlands}
}
\email{m.derijke@uva.nl}

\author{Jiafeng Guo}
\authornote{Corresponding author}
\orcid{0000-0002-9509-8674}
\affiliation{%
  \institution{Institute of Computing Technology, Chinese Academy of Sciences}
  \city{Beijing}
  \country{China}
}
\email{guojiafeng@ict.ac.cn}

\renewcommand{\shortauthors}{Xian et al.}

%% file: preamble/metadata.tex
\copyrightyear{2025}
\acmYear{2025}
\setcopyright{cc}
\setcctype{by}
\acmConference[SIGIR '25]{Proceedings of the 48th International ACM SIGIR
Conference on Research and Development in Information Retrieval}{July 13--18,
2025}{Padua, Italy}
\acmBooktitle{Proceedings of the 48th International ACM SIGIR Conference on
Research and Development in Information Retrieval (SIGIR '25), July 13--18,
2025, Padua, Italy}\acmDOI{10.1145/3726302.3729895}
\acmISBN{979-8-4007-1592-1/2025/07}

\ccsdesc[500]{Information systems~Evaluation of retrieval results}
\ccsdesc[500]{Information systems~Question answering}

\keywords{Long-form question answering, automatic evaluation}

%% file: tables/table_acurracy.tex
% asqa_main
\begin{table*}[!t]
\centering
\caption{Correlation coefficients between automatic metrics and human ratings for ASQA dataset. The best score for each column is highlighted in bold. The second best is underlined.}
\adjustbox{max width=\linewidth}{
\begin{tabular}{lccccccccc}
\toprule
\multirow{2}{*}{\makecell{Spearman/Kendall \\(\%)}}& \multicolumn{3}{c}{\textbf{Deterministic Metrics}} & \multicolumn{5}{c}{\textbf{Model-based Metrics}}\\
\cmidrule(r){2-4}
\cmidrule{5-10}
 & {RL} & {EM} & {DF1} & {BS} &{AR} & {FG} & {GPT-4o} & {Claude-3.5} & {Gemini-2.0} \\ 
\midrule
glm4-9b & 31.0/22.8 & 26.1/22.4 & 31.4/24.0 & 29.4/22.5 & 22.4/14.6 & \textbf{66.5}/\textbf{52.2} & \underline{56.6}/\underline{45.0} & 37.2/30.6 & 15.2/11.7 \\
gpt-3.5-turbo & -0.3/-0.5 & \underline{48.6}/\underline{40.3} & 18.5/12.9 & 2.6/2.4 & 8.3/5.7 & \textbf{50.8}/\textbf{40.6} & 32.6/25.1 & 22.3/18.3 & 15.4/11.9 \\
llama2-13b & 21.9/16.3 & \underline{56.6}/\underline{47.7} & 14.4/9.8 & 31.1/23.7 & 8.7/6.0 & \textbf{66.6}/\textbf{54.1} & 29.1/22.1 & 13.1/10.5 & 41.1/30.1 \\
llama2-7b & 17.0/12.1 & \textbf{54.3}/\textbf{46.2} & 14.7/11.1 & 22.9/16.6 & 23.7/17.3 & \underline{43.6}/\underline{35.6} & 32.5/24.6 & 30.8/22.6 & 22.6/15.4 \\
llama3-8b & -2.4/-1.6 & 47.5/39.4 & \underline{59.1}/\underline{44.8} & 12.1/8.1 & 14.0/10.1 & \textbf{72.1}/\textbf{60.5} & 43.5/33.8 & 20.7/17.5 & 32.1/23.8 \\
mistral-7b & 53.3/37.9 & 57.0/\textbf{49.9} & 33.2/24.0 & 33.5/24.0 & 17.1/12.0 & \textbf{62.0}/\underline{49.3} & 47.8/37.3 & \underline{59.0}/45.7 & 51.3/39.5 \\
solar-10.7b & 4.3/2.6 & 10.4/8.6 & -8.9/-6.1 & 9.0/8.4 & 8.9/7.0 & \underline{22.9}/\underline{22.3} & \textbf{41.4}/\textbf{32.2} & 20.6/17.7 & 8.5/6.8 \\
\midrule
Average & 11.4/8.0 & 41.4/\underline{34.9} & 23.0/16.8 & 16.2/11.7 & 12.2/8.7 & \textbf{55.0}/\textbf{44.9} & \underline{42.0}/32.4 & 33.0/26.3 & 32.4/24.4 \\
\bottomrule
\end{tabular}
}
\label{tab:asqa_main}
\end{table*}

\begin{table*}[h]
% todo: spearman, wikieval
\centering
\caption{Correlation coefficients between automatic metrics and human ratings for ANTIQUE and WikiEval dataset. The best score for each column is highlighted in bold. The second best is underlined.}
%\caption{Correlation coefficients between automatic metrics (\textit{RL} for Rouge-L, \textit{BERT} for BERTScore, \textit{A.R.} for Answer Relevance in RAGAS, \textit{CG} for coarse-grained GPT-4 scoring, \textit{FG} for fine-grained GPT-4 scoring) and human ratings for ANTIQUE dataset. The best score for each column is highlighted in bold. The second best is underlined.}
\begin{tabular}{p{2cm}cccccccccc}
\toprule
\multirow{2}{*}{\makecell{Spearman/Kendall \\(\%)}}& \multicolumn{5}{c}{\textbf{ANTIQUE dataset}} & \multicolumn{4}{c}{\textbf{WikiEval dataset}}\\ 
\cmidrule(r){2-6} \cmidrule{7-10}
& {RL} & {BS} &{AR} & {CG} & {FG} & {RL} & {BS} & {CG} & {FG}\\ 
\midrule
glm4-9b & -14.4/-11.4 & -9.8/-7.7 & 0.5/0.4 & \underline{36.7}/\underline{33.1} & \textbf{54.6}/\textbf{51.7} & \underline{44.4}/34.4 & \textbf{44.6}/34.4 & 38.1/\underline{34.8} & 39.1/\textbf{35.7} \\
gpt-3.5-turbo & 2.8/2.0 & -9.1/-7.0 & 43.7/35.0 & \underline{53.2}/\underline{47.1} & \textbf{56.5}/\textbf{51.7} & \textbf{57.1}/\textbf{45.1} & \underline{56.8}/\underline{43.1} & 41.9/36.5 & 46.9/42.1 \\
llama2-13b & 11.0/7.8 & 0.5/0.5 & 10.9/7.6 & \underline{74.9}/\underline{62.5} & \textbf{81.5}/\textbf{71.2} & \underline{61.4}/\underline{47.2} & 57.1/44.3 & \textbf{61.8}/\textbf{51.7} & 55.4/46.8 \\
llama2-7b & 13.8/10.0 & 9.3/6.9 & 17.7/13.4 & \underline{69.3}/\underline{58.6} & \textbf{74.6}/\textbf{64.7} & \underline{74.6}/\textbf{60.4} & \textbf{75.2}/\underline{60.3} & 60.8/52.8 & 60.1/52.9 \\
llama3-8b & 2.4/1.5 & 8.7/6.3 & 2.5/1.6 & \underline{65.6}/\underline{54.6} & \textbf{72.2}/\textbf{63.1} & \textbf{51.7}/\textbf{40.6} & \underline{44.9}/34.5 & 44.1/\underline{37.7} & 35.7/31.9 \\
mistral-7b & 29.2/21.0 & 25.9/18.5 & 23.3/16.9 & \underline{73.0}/\underline{62.5} & \textbf{83.9}/\textbf{73.6} & \underline{51.3}/40.0 & \textbf{60.7}/\textbf{48.9} & 47.7/40.0 & 50.4/\underline{42.6} \\
solar-10.7b & -4.6/-3.4 & -9.7/-6.6 & -3.5/-2.3 & \underline{54.3}/\underline{46.2} & \textbf{70.8}/\textbf{62.5} & \textbf{56.3}/\textbf{43.3} & \underline{54.3}/\underline{41.7} & 41.2/35.5 & 34.1/31.1 \\
\midrule
Average & -7.9/-5.9 & -24.4/-17.2 & 28.3/20.6 & \underline{83.0}/\underline{70.9} & \textbf{85.2}/\textbf{75.6} & \textbf{55.8}/\underline{43.4} & \underline{55.5}/42.6 & 51.4/\textbf{43.9} & 49.0/43.1 \\
\bottomrule
\end{tabular}
\label{tab:antique_main}
\end{table*}

%% file: tables/table_temperatures.tex
\begin{table*}[h]
\caption{The relationship between model scores and evaluators' temperature, with the default temperature set to 0.}
\label{temperature_analysis}
\centering
\begin{tabular}{l cc ccc ccc cc ccc ccc}
\toprule
& \multicolumn{8}{c}{\textbf{ASQA}}  & \multicolumn{8}{c}{\textbf{Wikieval}} \\
\cmidrule(r){2-9}
\cmidrule{10-17}
& {\textbf{Scores}} & {\textbf{Ranks}} & \multicolumn{3}{c}{\textbf{Scores Gain}} & \multicolumn{3}{c}{\textbf{Rank Gain}} & {\textbf{Scores}} & {\textbf{Ranks}} & \multicolumn{3}{c}{\textbf{Scores Gain}} & \multicolumn{3}{c}{\textbf{Rank Gain}} \\
\cmidrule(r){2-3}
\cmidrule(r){4-6}
\cmidrule(r){7-9}
\cmidrule(r){10-11}
\cmidrule(r){12-14}
\cmidrule{15-17}
 &\textbf{0.0} &\textbf{0.0} & \textbf{0.3} & \textbf{0.7} & \textbf{1.0} 
 &  \textbf{0.3} & \textbf{0.7} & \textbf{1.0}  &\textbf{0.0} &\textbf{0.0} & \textbf{0.3} & \textbf{0.7} & \textbf{1.0} 
 &  \textbf{0.3} & \textbf{0.7} & \textbf{1.0} \\
\midrule

\rowcolor{gray!10}\textbf{claude-3.5} & 8.66 & 1 & \cellcolor{red!12.00}{0.08} & \cellcolor{red!6.90}{0.05} & \cellcolor{blue!6.60}{-0.04} & {0.0} & {0.0} & {0.0} & 8.26 & 2 & \cellcolor{blue!21.00}{-0.14} & \cellcolor{blue!6.00}{-0.04} & \cellcolor{blue!12.00}{-0.08} & \cellcolor{blue!25.00}{-1.0} & \cellcolor{blue!25.00}{-1.0} & \cellcolor{blue!25.00}{-1.0} \\
\textbf{gpt-4o} & 8.13 & 2 & \cellcolor{red!6.90}{0.05} & \cellcolor{red!5.70}{0.04} & \cellcolor{red!5.40}{0.04} & {0.0} & {0.0} & {0.0} & 8.08 & 4 & \cellcolor{blue!27.00}{-0.18} & \cellcolor{red!36.00}{0.24} & \cellcolor{red!24.00}{0.16} & \cellcolor{blue!50.00}{-2.0} & \cellcolor{red!50.00}{2.0} & \cellcolor{red!50.00}{2.0} \\
\rowcolor{gray!10}\textbf{gemini-2.0} & 6.69 & 3 & \cellcolor{red!6.00}{0.04} & \cellcolor{red!3.90}{0.03} & \cellcolor{red!14.10}{0.09} & {0.0} & {0.0} & {0.0} & 7.86 & 7 & \cellcolor{blue!60.00}{-0.40} & \cellcolor{blue!42.00}{-0.28} & \cellcolor{blue!12.00}{-0.08} & \cellcolor{blue!25.00}{-1.0} & \cellcolor{blue!25.00}{-1.0} & \cellcolor{blue!25.00}{-1.0} \\
\textbf{glm4-9b} & 4.87 & 4 & \cellcolor{blue!2.70}{-0.02} & \cellcolor{red!0.30}{0.00} & \cellcolor{blue!0.90}{-0.01} & {0.0} & {0.0} & {0.0} & 8.62 & 1 & \cellcolor{blue!24.00}{-0.16} & \cellcolor{blue!12.00}{-0.08} & \cellcolor{blue!30.00}{-0.20} & {0.0} & {0.0} & {0.0} \\
\rowcolor{gray!10}\textbf{solar-10.7b} & 4.70 & 5 & \cellcolor{red!6.90}{0.05} & \cellcolor{red!9.60}{0.06} & \cellcolor{red!9.90}{0.07} & {0.0} & {0.0} & {0.0} & 7.94 & 5 & \cellcolor{red!24.00}{0.16} & \cellcolor{red!33.00}{0.22} & \cellcolor{red!27.00}{0.18} & \cellcolor{red!25.00}{1.0} & \cellcolor{red!12.50}{0.5} & \cellcolor{red!25.00}{1.0} \\
\textbf{gpt-3.5-turbo} & 4.67 & 6 & \cellcolor{red!1.80}{0.01} & \cellcolor{red!6.90}{0.05} & \cellcolor{blue!6.90}{-0.05} & {0.0} & {0.0} & {0.0} & 7.92 & 6 & \cellcolor{red!12.00}{0.08} & \cellcolor{blue!9.00}{-0.06} & \cellcolor{red!3.00}{0.02} & \cellcolor{red!25.00}{1.0} & {0.0} & {0.0} \\
\rowcolor{gray!10}\textbf{llama3-8b} & 4.27 & 7 & \cellcolor{red!15.30}{0.10} & \cellcolor{red!13.50}{0.09} & \cellcolor{red!13.80}{0.09} & {0.0} & {0.0} & {0.0} & 7.66 & 8 & \cellcolor{blue!3.00}{-0.02} & \cellcolor{red!27.00}{0.18} & \cellcolor{red!24.00}{0.16} & \cellcolor{red!25.00}{1.0} & \cellcolor{red!25.00}{1.0} & \cellcolor{red!25.00}{1.0} \\
\textbf{llama2-13b} & 4.01 & 8 & \cellcolor{blue!3.60}{-0.02} & \cellcolor{red!8.10}{0.05} & \cellcolor{blue!8.70}{-0.06} & {0.0} & {0.0} & {0.0} & 6.82 & 9 & \cellcolor{red!6.00}{0.04} & \cellcolor{red!12.00}{0.08} & \cellcolor{red!27.00}{0.18} & {0.0} & {0.0} & {0.0} \\
\rowcolor{gray!10}\textbf{llama2-7b} & 3.60 & 9 & \cellcolor{blue!9.30}{-0.06} & \cellcolor{blue!3.30}{-0.02} & \cellcolor{blue!2.70}{-0.02} & {0.0} & {0.0} & {0.0} & 6.60 & 10 & \cellcolor{red!6.00}{0.04} & \cellcolor{red!21.00}{0.14} & \cellcolor{blue!6.00}{-0.04} & {0.0} & {0.0} & {0.0} \\
\textbf{mistral-7b} & 3.03 & 10 & \cellcolor{red!4.50}{0.03} & \cellcolor{red!3.30}{0.02} & \cellcolor{red!3.30}{0.02} & {0.0} & {0.0} & {0.0} & 8.20 & 3 & \cellcolor{blue!6.00}{-0.04} & \cellcolor{blue!6.00}{-0.04} & \cellcolor{blue!33.00}{-0.22} & \cellcolor{red!25.00}{1.0} & \cellcolor{blue!37.50}{-1.5} & \cellcolor{blue!50.00}{-2.0} \\

\bottomrule
\end{tabular}
\end{table*}

%% file: figures/fairness_length.tex
\begin{figure}
    \centering
        
    \begin{subfigure}{\columnwidth}
     \includegraphics[width=\columnwidth]{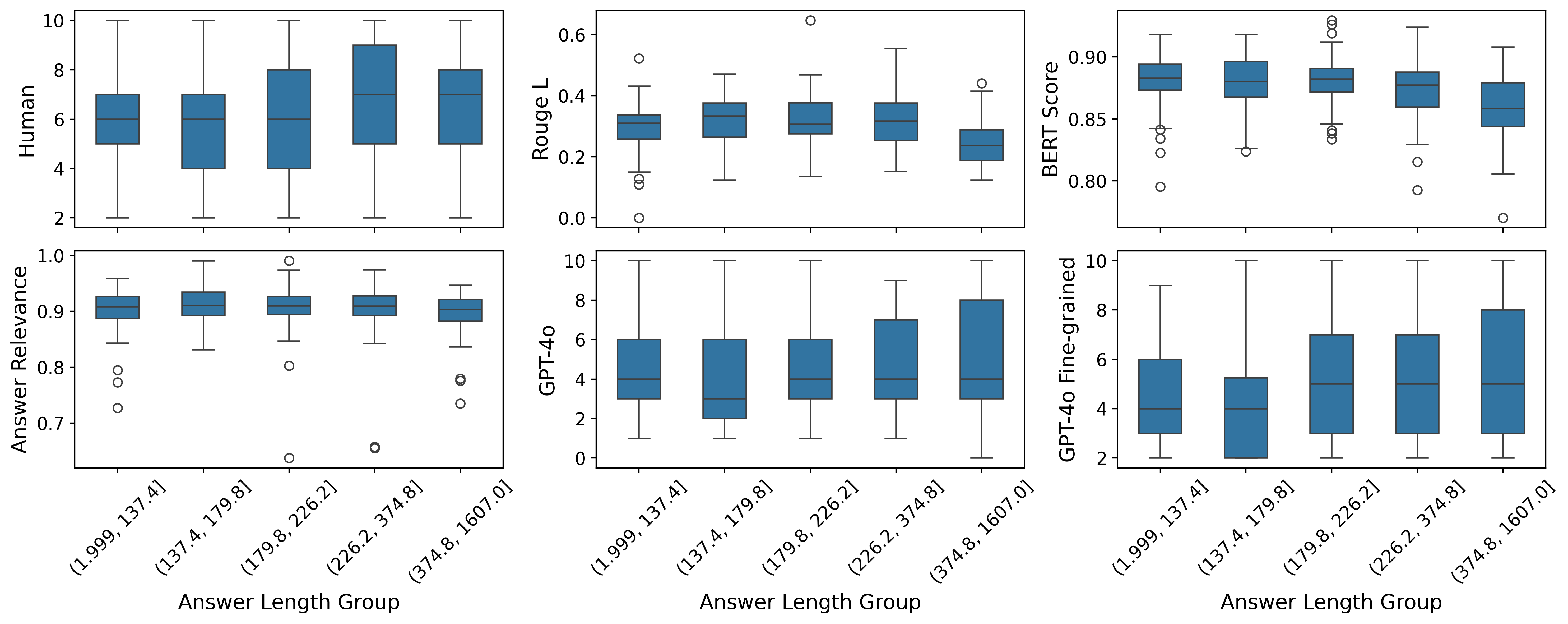}
    \caption{Results on the ASQA dataset}
    \label{fig:length_box_asqa}   
    \end{subfigure}

    % \vspace{0.2cm}
    
    \begin{subfigure}{\columnwidth}
    \includegraphics[width=\columnwidth]{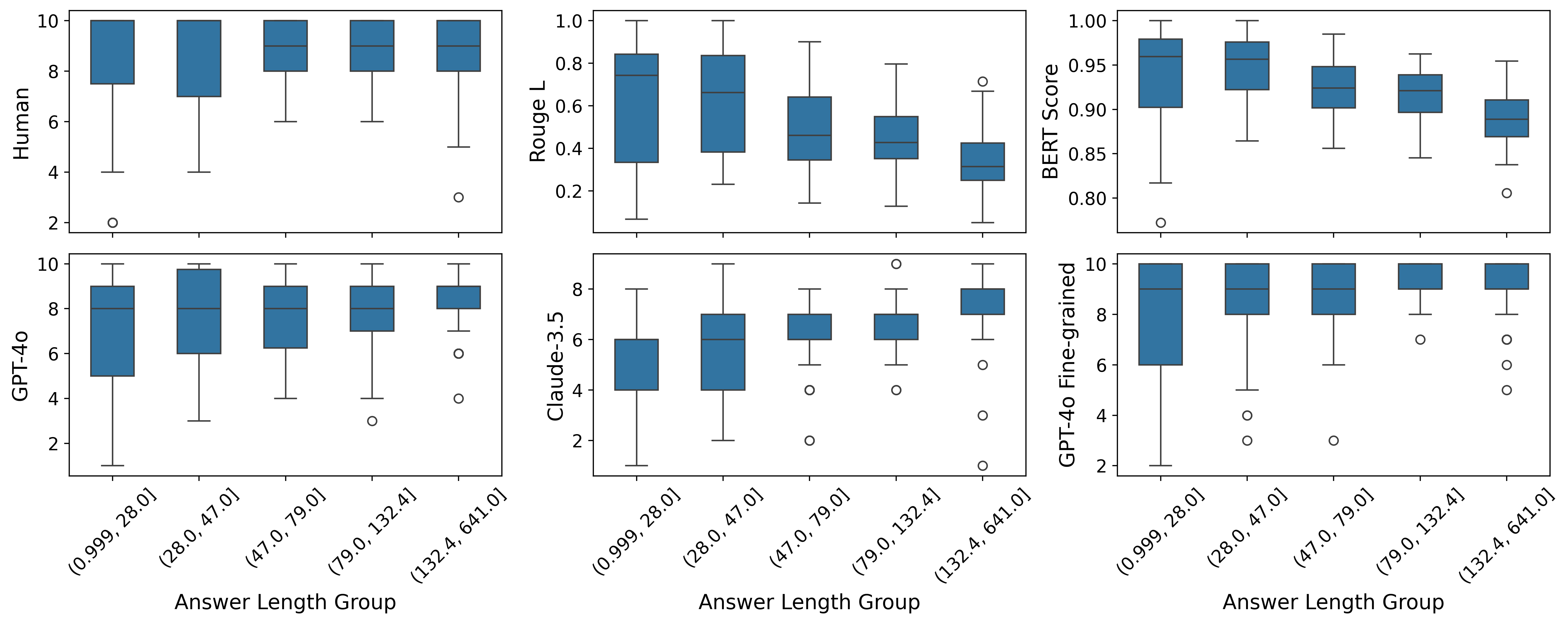}
    \caption{Results on the Wikieval dataset}
    \label{fig:length_box_wikieval}
    \end{subfigure}
    
    % \vspace{0.2cm}
    
    \begin{subfigure}{\columnwidth}
    \includegraphics[width=\columnwidth]{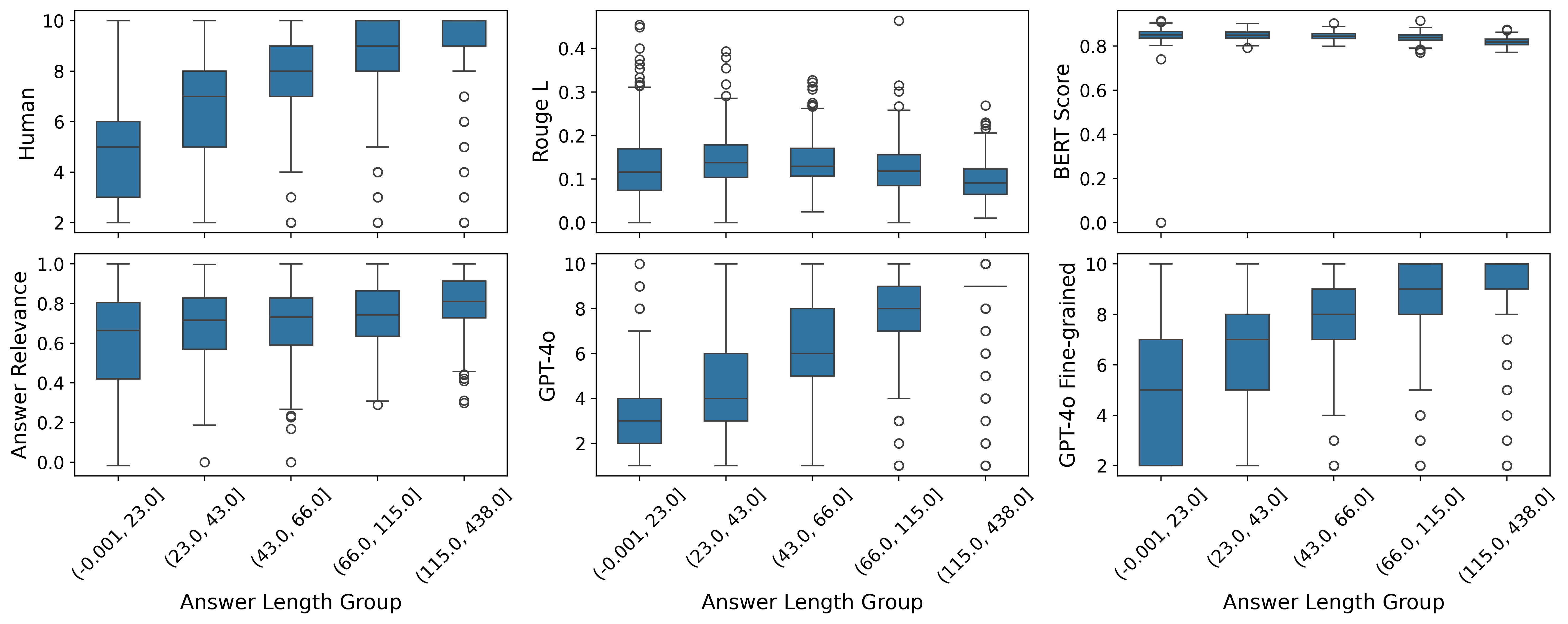}
    \caption{Results on the Antique dataset}
    \label{fig:length_box_antique}  
    \end{subfigure}
    \caption{Relationship between answer length and metrics.}
    \label{fig:length_box}
\end{figure}

%% file: figures/self-reinforcing-heatmap.tex
\begin{figure}[!t]
    \centering
    \begin{tikzpicture}[scale=0.68]
        \begin{groupplot}[
            group style={
                group size=1 by 3,
                horizontal sep=0.2cm,
                vertical sep=0.2cm
            },
            colormap={coolwarm}{
                rgb255(0cm)=(58,76,192);     % 深蓝色
                rgb255(0.25cm)=(139, 174, 253)
                rgb255(0.5cm)=(220, 220, 221); % 白色
                rgb255(0.75cm)=(244, 154, 123)
                rgb255(1cm)=(179, 3, 38)       % 深红色
            },
            colorbar style={
                at={(1.05,3)},
                height=8cm,
                width=0.3cm,
                ytick=\empty
            },
            ytick={0,...,2},
            yticklabels={gemini-2.0,claude-3.5,gpt-4o},
            y tick label style={anchor=east, font=\small},
            width=.6\textwidth,
            height=.25\textwidth,
            point meta min=0,
            point meta max=1,
            mesh/cols=10,
            enlarge x limits=false,
            enlarge y limits=false,
            nodes near coords,
            nodes near coords align={center},
            every node near coord/.append style={
                font=\footnotesize,
                text=black,
                anchor=center
            },
            nodes near coords style={
                /pgf/number format/fixed,
                /pgf/number format/precision=3
            }
        ]
        \nextgroupplot[ ylabel={GPT-4o}, ylabel style={yshift=10pt}, xticklabels=\empty]
            \addplot[matrix plot*,point meta=explicit] table[meta=meta] {
            x y meta
            0 2 0.500
            1 2 0.337
            2 2 0.738
            3 2 0.920
            4 2 0.851
            5 2 0.882
            6 2 0.924
            7 2 0.941
            8 2 0.953
            9 2 0.986
            0 1 0.663
            1 1 0.500
            2 1 0.837
            3 1 0.937
            4 1 0.908
            5 1 0.922
            6 1 0.956
            7 1 0.965
            8 1 0.962
            9 1 0.978
            0 0 0.262
            1 0 0.163
            2 0 0.500
            3 0 0.796
            4 0 0.725
            5 0 0.758
            6 0 0.840
            7 0 0.860
            8 0 0.878
            9 0 0.931
            };

        \nextgroupplot[ ylabel={Claude-3.5}, ylabel style={yshift=10pt}, xticklabels=\empty]
        \addplot[matrix plot*,point meta=explicit] table[meta=meta] {
            x y meta
            0 2 0.500
            1 2 0.192
            2 2 0.744
            3 2 0.897
            4 2 0.792
            5 2 0.777
            6 2 0.921
            7 2 0.940
            8 2 0.931
            9 2 0.965
            0 1 0.808
            1 1 0.500
            2 1 0.940
            3 1 0.980
            4 1 0.948
            5 1 0.938
            6 1 0.986
            7 1 0.988
            8 1 0.988
            9 1 0.989
            0 0 0.256
            1 0 0.060
            2 0 0.500
            3 0 0.764
            4 0 0.656
            5 0 0.626
            6 0 0.835
            7 0 0.867
            8 0 0.895
            9 0 0.935
            };

        % 第三个子图
        \nextgroupplot[
            ylabel={Gemini-2.0},
            ylabel style={yshift=10pt},
            xtick={0,...,9},
            x tick label style={rotate=45, anchor=east, font=\small, xshift=-2pt},
            xticklabels={gpt-4o, claude-3.5,gemini-2.0,gpt-3.5-turbo,glm4-9b,solar-10.7b,llama3-8b,llama2-13b,llama2-7b,mistral-7b},
            colorbar
        ]
        \addplot[matrix plot*,point meta=explicit] table[meta=meta] {
            x y meta
            0 2 0.500
            1 2 0.413
            2 2 0.532
            3 2 0.785
            4 2 0.716
            5 2 0.701
            6 2 0.778
            7 2 0.804
            8 2 0.857
            9 2 0.901
            0 1 0.587
            1 1 0.500
            2 1 0.612
            3 1 0.819
            4 1 0.774
            5 1 0.749
            6 1 0.815
            7 1 0.848
            8 1 0.882
            9 1 0.916
            0 0 0.468
            1 0 0.388
            2 0 0.500
            3 0 0.761
            4 0 0.681
            5 0 0.673
            6 0 0.758
            7 0 0.783
            8 0 0.844
            9 0 0.887
            };
    \end{groupplot}
    
    \end{tikzpicture}
    \caption{Winrate for different metrics on the ASQA dataset.}
    \label{fig:self_reinforce}
\end{figure}
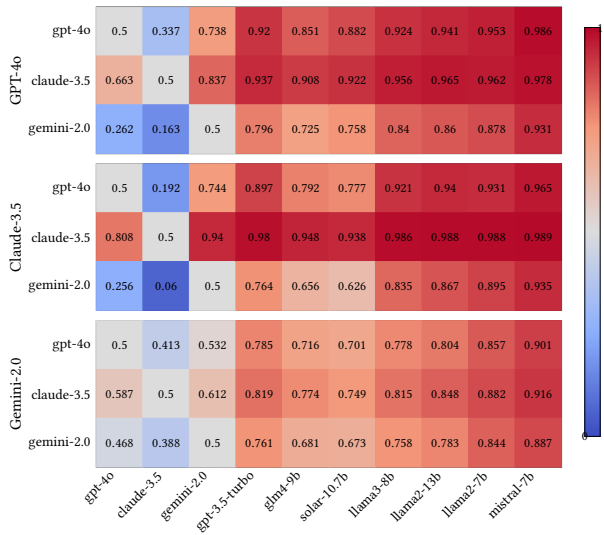

%% file: tables/table_enhance.tex
\begin{table*}[!htbp]
\centering
\caption{Variation of correlation coefficients under different prompt words.}
\label{tab:model_performance}
\begin{tabular}{l ccc ccccccccc}
\toprule
\multirow{2}{*}{\makecell{Kendall (\%)}} &\multicolumn{3}{c}{LLMs} &\multicolumn{9}{c}{Prompts}\\
\cmidrule(r){2-4}
\cmidrule{5-13}
 & {GPT-4o} & {Claude-3.5} & {Gemini-2.0} & {P1} & {P2} & {P3} & {P4} & {P5} & {P6} & {P7} & {P8} & {P9} \\ 
\midrule
\rowcolor{gray!10}\textbf{glm4-9b} & 34.76 & \cellcolor{blue!54.23}{-36.15} & \cellcolor{blue!4.08}{-2.72} & \cellcolor{red!12.85}{8.56} & \cellcolor{blue!26.17}{-17.45} & \cellcolor{red!5.22}{3.48} & \cellcolor{red!14.81}{9.87} & \cellcolor{red!5.87}{3.91} & \cellcolor{red!6.15}{4.10} & \cellcolor{red!3.38}{2.25} & \cellcolor{blue!3.57}{-2.38} & \cellcolor{red!11.23}{7.49} \\
\textbf{gpt-3.5-turbo} & 36.47 & \cellcolor{blue!3.03}{-2.02} & \cellcolor{blue!9.72}{-6.48} & \cellcolor{red!24.66}{16.44} & \cellcolor{red!0.92}{0.61} & \cellcolor{red!41.46}{27.64} & \cellcolor{red!33.80}{22.54} & \cellcolor{red!12.53}{8.36} & \cellcolor{red!22.65}{15.10} & \cellcolor{red!18.53}{12.36} & \cellcolor{red!19.86}{13.24} & \cellcolor{red!29.76}{19.84} \\
\rowcolor{gray!10}\textbf{llama2-13b} & 51.66 & \cellcolor{blue!35.25}{-23.50} & \cellcolor{red!7.39}{4.93} & \cellcolor{red!24.80}{16.54} & \cellcolor{red!15.79}{10.52} & \cellcolor{red!17.60}{11.73} & \cellcolor{red!12.52}{8.35} & \cellcolor{red!16.98}{11.32} & \cellcolor{red!18.16}{12.11} & \cellcolor{red!18.19}{12.13} & \cellcolor{red!18.61}{12.41} & \cellcolor{red!12.93}{8.62} \\
\textbf{llama2-7b} & 52.77 & \cellcolor{blue!16.75}{-11.17} & \cellcolor{blue!7.32}{-4.88} & \cellcolor{red!9.12}{6.08} & \cellcolor{red!4.28}{2.85} & \cellcolor{red!11.68}{7.78} & \cellcolor{red!15.32}{10.22} & \cellcolor{red!6.64}{4.43} & \cellcolor{red!8.54}{5.69} & \cellcolor{red!0.08}{0.05} & \cellcolor{red!9.03}{6.02} & \cellcolor{red!8.68}{5.79} \\
\rowcolor{gray!10}\textbf{llama3-8b} & 37.66 & \cellcolor{blue!2.55}{-1.70} & \cellcolor{blue!27.37}{-18.25} & \cellcolor{red!28.39}{18.93} & \cellcolor{red!27.36}{18.24} & \cellcolor{red!11.53}{7.69} & \cellcolor{red!32.59}{21.73} & \cellcolor{red!23.00}{15.33} & \cellcolor{red!5.45}{3.63} & \cellcolor{red!25.83}{17.22} & \cellcolor{red!23.85}{15.90} & \cellcolor{red!23.51}{15.67} \\
\textbf{mistral-7b} & 39.97 & \cellcolor{blue!47.13}{-31.42} & \cellcolor{blue!46.59}{-31.06} & \cellcolor{blue!9.66}{-6.44} & \cellcolor{blue!15.09}{-10.06} & \cellcolor{blue!10.17}{-6.78} & \cellcolor{blue!24.28}{-16.19} & \cellcolor{blue!25.95}{-17.30} & \cellcolor{blue!23.81}{-15.88} & \cellcolor{blue!10.10}{-6.74} & \cellcolor{blue!24.86}{-16.58} & \cellcolor{blue!21.28}{-14.19} \\
\rowcolor{gray!10}\textbf{solar-10.7b} & 35.45 & \cellcolor{blue!12.23}{-8.16} & \cellcolor{red!12.83}{8.56} & \cellcolor{red!18.44}{12.29} & \cellcolor{red!22.51}{15.01} & \cellcolor{red!32.59}{21.72} & \cellcolor{red!22.19}{14.79} & \cellcolor{red!14.83}{9.89} & \cellcolor{red!6.98}{4.65} & \cellcolor{red!29.91}{19.94} & \cellcolor{red!21.92}{14.61} & \cellcolor{red!36.15}{24.10} \\
\midrule
\textbf{Average} & 43.89 & \cellcolor{blue!23.87}{-15.91} & \cellcolor{blue!7.26}{-4.84} & \cellcolor{red!17.52}{11.68} & \cellcolor{red!8.22}{5.48} & \cellcolor{red!18.66}{12.44} & \cellcolor{red!16.73}{11.15} & \cellcolor{red!11.56}{7.71} & \cellcolor{red!8.61}{5.74} & \cellcolor{red!14.23}{9.49} & \cellcolor{red!12.32}{8.21} & \cellcolor{red!17.07}{11.38} \\
\bottomrule
\end{tabular}
\label{tab:enhance}
\end{table*}

%% file: tables/table_prompt_settings.tex
\begin{table}
    \caption{Prompt settings.}
    \label{tab:prompts}
    \centering
    \begin{tabular}{cl}
    \toprule
       \textbf{Prompt}  & \textbf{Settings} \\
    \midrule
        P1 & Task+Data+Output+Criteria \\
        P2 & Task+Data+Output \\
        P3 & Task+Data+Criteria+Output \\
        P4 & Output+Task+Data+Criteria \\
        P5 & Task+Output+Criteria \\
        P6 & Data+Output+Criteria \\
        P7 & Output+Criteria+Task+Data \\
        P8 & Data+Task+Criteria(short)+Output \\
        P9 & Data+Task+Criteria+Output \\
    \bottomrule
    \end{tabular}
\end{table}

%% file: tables/table_instruction.tex
\begin{table*}[!ht]
    \centering
    \caption{Evaluation instructions.}
    \setlength{\tabcolsep}{.5mm}
    \begin{tabularx}{\textwidth}{lX}
    \toprule
    \multicolumn{1}{c}{\textbf{Components}} & 
    \multicolumn{1}{c}{\textbf{Instructions}} \\
    \midrule
    Task & Score the following LLM output of a question-answering task with respect to the following aspects using a 1 to 5 star rating system. \\
    \midrule
    Data & The dataset is a Factoid Question-Answering dataset, specifically designed for evaluating factual precision and detailed comparative reasoning in AI-generated answers. \\
    \midrule
    Output & Begin your evaluation by providing a short explanation. Be as objective as possible. After providing your explanation, please provide your evaluation by strictly following the JSON format, such as: [[SCORE]] \{"accuracy": 2, "informativeness": 3\}. \\
    \midrule
    Criteria & \makecell[l]{
        \textbf{Accuracy}: Determine the accuracy of the answers, verifying the correctness and reliability of the information provided. \\
        1 star: Incorrect information \\
        2 stars: Partially correct information \\
        3 stars: Half correct information \\
        4 stars: Mostly correct information \\
        5 stars: Perfectly correct information \\
        \textbf{Informativeness}: Examines whether the answers provide sufficient and meaningful information that is useful to \\the user and relevant to the question. \\
        1 star: No information or irrelevant information \\
        2 stars: Very little information \\
        3 stars: Some information \\
        4 stars: Enough information \\
        5 stars: Highly informative
    } 
    \\
    \bottomrule
    \end{tabularx}
    \label{tab:instructions}
\end{table*}